\documentclass{article}
\usepackage{graphicx}
\usepackage{makeidx}
\usepackage{amsmath}
\usepackage{amsthm}
\usepackage{amsfonts}
\usepackage{amssymb}
\usepackage{epsfig}
\usepackage{tabularx}
\usepackage{latexsym}
\usepackage[english]{babel}
\usepackage{verbatim}

\newtheorem{conjec}{Conjecture}

\newcommand{\noi}{\noindent}
\newcommand{\NC}{\mathcal{N}}
\newcommand{\LA}{\mathcal{L}}
\newcommand{\SM}{\mathcal{S}}
\newcommand{\ep}{\epsilon}
\newcommand{\defn}{\stackrel{def}{=}}

\begin{document}
\title{On the Rise and Fall of Online Social Networks}

\author{
	Arnab Basu
        \footnote{Bilkent University, Ankara, Turkey. Email: arnab@bilkent.edu.tr}
        \footnote{Corresponding Author}
        \and
        Saswata Shannigrahi 
       \footnote{Saint Petersburg State University, Russia. Email: saswata.shannigrahi@gmail.com}
        \footnote{Corresponding Author}
        \and
        Simrat Singh Chhabra 
        \footnote{University of Southern California, USA. Email: simratsc@usc.edu}
        \and
	Ajit Brundavanam
        \footnote{Indian Institute of Technology Kharagpur, India. Email: ajitb@iitkgp.ac.in}
}
\maketitle

\begin{abstract}
\noi The rise and fall of online social networks recently generated an enormous amount of interest among people, both inside and outside of academia. Gillette [Businessweek magazine, 2011] did a detailed analysis of MySpace, which started losing its popularity since 2008. Cannarella and Spechler [ArXiv, 2014] used a model of disease spread to explain the rise and fall of MySpace. In this paper, we present a graph theoretical model that may be able to provide an alternative explanation for the rise and fall of online social networks. Our model is motivated by the well-known Barab\'{a}si-Albert model of generating random scale-free networks using preferential attachment or `rich-gets-richer' phenomenon. As shown by our empirical analysis, we conjecture that such an online social network growth model is inherently flawed as it fails to maintain the stability of such networks while ensuring their growth. In the process, we also conjecture that our model of preferential attachment also exhibits scale-free phenomenon. \end{abstract}

\section{Introduction}
\noindent The last few decades have seen the emergence of highly popular online social networks like MySpace, Orkut and Facebook. MySpace was founded in 2003 and it gained its peak popularity in 2008 \cite{gillette}. However, most of the users started abandoning it since 2008 \cite{cannarella, gillette}. Cannarella and Spechler \cite{cannarella} modified the SIR model of disease spread (see, e.g., \cite{bailey}) to explain this phenomenon. The disease spread model has also been used to study the arrival and departure dynamics of the users in online social networks \cite{wu}. 

In this paper, we propose a model that may be able to provide an alternative explanation for the rise and fall of online social networks. To begin with, we note that people join online social networks due to the presence of their friends in these networks and leave due to the inactivity of their friends \cite{wu}. It is also known that a large percentage of online social network friendships are not active or strong friendships between the users \cite{easley, huberman, marlow}, where a {\it strong friendship} between a pair of online social network friends is indicated by regular communications between them. In fact, the well-known Dunbar's number \cite{dunbar} says that an individual can comfortably maintain stable relationships with at most $150$ other people only. It means that the number of strong friendships of a person is likely to be limited to $150$ in any social network, even though there is no bound on the number of friends one can have in most online social networks. 

In order to identify strong friendships between people in a mobile communication network, Onella et. al \cite{onnela} did an empirical study to establish that a greater neighborhood overlap between a pair of friends corresponds to a stronger friendship between them. We assume that such a result is also true for online social networks. The {\it neighborhood overlap} between a pair of friends $A$ and $B$ in such a network (represented as an undirected graph) is defined as the number of users that are neighbors of both $A$ and $B$ divided by the number of users that are neighbors of either of them (except $A$ and $B$ themselves) \cite{easley}. We define a pair of friends to be {\it strong friends} if their neighborhood overlap is at least a given constant. 

In this paper, we study the change in the size of the largest connected component (LCC) in the strong friendship subgraph of an evolving random graph defined below in Section \ref{model-sec}. This is motivated by the observation that the nodes in the core of an online social network are more likely to survive than the nodes at the periphery \cite{wu}. We consider the LCC in the strong friendship subgraph of an online social network to be its {\it core} that is important to retain most of its users.

\section{Model} \label{model-sec}

Our model is motivated by the well-known Barab\'{a}si-Albert (BA) model \cite{barabasi} that can be used to generate random scale-free networks based on preferential attachment and growth. The preferential attachment is the property that a node with  higher degree than another node is more likely than the other node to get connected to new nodes as the network grows, i.e., a manifestation of the `rich-gets-richer' phenomenon.

Starting with an initial collection of $m_0 > 0$ nodes where $m_0$ is a given natural number, the BA model adds one node at a time to the network. In each time step, a new node gets connected to $m \leq m_0$ existing nodes. The probability $p_i(t)$ that the new node at time $t + 1$ is connected to an existing node $v_i$ is 
\begin{equation} \label{BAprob}
p_i(t) = \frac{d_i(t)}{\sum_j d_j(t)}, 
\end{equation}
where $d_i(t)$ is the degree of node $v_i$ at time $t$ and $\sum_j d_j(t)$ is the corresponding sum of the degrees of all nodes in the current network. We assume that the nodes $v_{m_0 + 1}, v_{m_0 + 2}, \ldots, v_{m_0 +t}, \ldots$ are ordered in the sequence of their additions to the network, following an arbitrary given ordering $<v_1, v_2, \ldots, v_{m_0}>$ of the initial $m_0$ nodes. 

Let $G_0$ be the initial (time $t=0$) graph with $m_0$ nodes, and $G_t \defn (V_t,E_t)$ be the random graph with the set $V_t$ of $m_t = m_0 + t$ nodes after $t$ nodes have been added along with the addition of $mt$ edges $\{e_1,\ldots,e_{mt}\}$ by the BA model, i.e.,
$E_t = \{e_1,\ldots,e_{mt}\} \cup E_0$ where $E_0$ is the initial set of edges between the starting $m_0$ nodes. For a given $\epsilon > 0$, we define two friends $v_i,v_j \in V_t$ to be strong friends at time $t$ if their neighborhood overlap $\NC_t(v_i,v_j)$ at time $t$ is greater than or equal to $\ep$, i.e., 
\begin{equation} \label{neighbour}
	\NC_t(v_i,v_j) \defn \frac{|\{v' \in V_t : e_{v_i v'} = e_{v_j v'} =1\}|}{d_i(t) + d_j(t) - 2- |\{v' \in V_t : e_{v_i v'} = e_{v_j v'} =1\}} \geq \ep,
\end{equation}
where, for a pair of nodes $v,v'$, $e_{vv'} = 1$ implies the existence of an edge between them and $0$ implying otherwise.

 Given $\ep > 0$, the time-$t$ strong friendship subgraph $G^s_t \subseteq G_t$ is defined as follows: $G^s_t$ contains the same set of nodes as that of $G_t$, and a pair of nodes $i$ and $j$ are connected by an edge in $G^s_t$ if and only if they are connected by an edge in $G_t$ and their neighborhood overlap in $G_t$ is greater than or equal to $\ep$, i.e., $G^s_t \defn (V_t, E^s_t)$ where $E^s_t \defn \{\{v_i,v_j\} \in E_t : \NC_t(v_i,v_j) \geq \ep\}$. At time $t$, the corresponding LCC in $G^s_t$ is denoted by $\LA_t$. For any set $S$, $\|S\|$ denotes its cardinality. 

Before proceeding further, we would like to make some clarifications. As is well-noted in the literature (see, e.g., \cite[Chap.\ 8]{HFSTD01} and references therein), the description of the BA model as originally proposed in \cite{barabasi} is quite imprecise and does not explain several basic facts, namely, how the first edge is connected, what the dependencies are between the $m$ edges added at any given time $t$ and, more importantly, how to address the issue that the expected number of edges connected a new incoming vertex at any time $t$ to earlier existing vertices is not $m$ (as the BA model claims it to be) but actually $1$.

To address these issues, we use the following algorithm which we henceforth refer to as the {\it PArallel Preferential Attachment} or the PAPA Algorithm or simply PAPA. It works as follows: To start with, it mandates that $G_0$ with $m_0$ nodes has some given nonzero degrees $d_i(0),\ i=1,\ldots,m_0$. If $m_0 = 1$, then we initialize $d_1(0) = 1$ as a boundary case so that, naturally as required, the new (second) node coming in will surely ($p_1(0) = 1$) get connected to this initial existing node, i.e., exactly $m = m_0 = 1$ edge will be created between these two nodes. In general, we use the following iterative setup to simultaneously (in parallel) add the $m \leq m_0$ edges from a new node $v_{m_0+t+1}$ at time $t+1$ to the $m_0 + t$ nodes existing till time $t$. At any time step $t+1$, PAPA runs as follows:
\begin{enumerate}
\item We start with a set $\SM \equiv \emptyset$.
\item The interval $[0, 1]$ is first divided into $m_0 + t$ non-overlapping sub-intervals $[p_0(t) \equiv 0, p_1(t)), [p_1(t), p_1(t)+p_2(t)), \ldots, [\sum_{i=0}^{m_0 + t-1} p_i(t), \sum_{i=0}^{m_0 + t} p_i(t) \equiv 1]$ where $p_i(t)$ is given by (\ref{BAprob}) above.
\item Now, a random number $X \sim U[0,1]$ is generated and suppose it falls in the interval 
$[\sum_{i=0}^{l-1} p_i(t), \sum_{i=0}^{l} p_i(t))$ for some $1 \leq l \leq m_0+t$. 
\item Then, we update $\SM \leftarrow \SM \cup \{l\}$ and delete this interval of length $p_l(t)$ from $[0,1]$ subsequently.
\item Now, we stretch each of the remaining intervals by multiplying with $\frac{1}{1-p_l(t)}$.
\item If $m>1$, the next update to $\SM$ is done in the same way using these new re-scaled intervals for each subsequent edge. These steps are repeated from Step 3 above until all $m$ updates are made.
\item Finally, for each $l \in \SM$, we add an edge $\{v_{m_0+t+1},v_l\}$ to the current configuration of $G_t$, i.e., make $e_{\{v_{m_0+t+1} v_l\}} = 1$.
\end{enumerate}

We refer to the graph generated by this Algorithm as the {\it PAPA Random Graph} and note three important points below about this Algorithm:
\begin{itemize}
	\item Unlike most of the existing literature in this domain 
	(see, again, \cite[Chap.\ 8]{HFSTD01} and references therein), PAPA 
	eliminates any possibility of multi-edges and self-loops as is obvious from Step 4 above. This property is crucial for real-life online networks where there is either a real online connection between two persons or there is none, and, physical justification of multiple connections between two persons is unrealistic. Since our primary aim is to study online social networks, we adopt this approach in this paper.
This approach has also been adopted in \cite[Page\ 3]{berger} and \cite[Page\ 150]{garavaglia}.
	\item The second important point is that we do not re-normalize our $p_i(t)$-s during the addition of the $m$ edges at any time point $t+1$, i.e., we hold those values constant at time $t+1$; hence, it does not reflect the inter-dependencies of the $m$ edges being added at that time $t+1$ through probability re-scaling. This actually provides an exact parallel or simultaneous preferential attachment scheme as is realistically done by a new user coming in to join an existing online social network. 
	\item Finally, as is clear from Step 6 above, we add exactly $m$ edges at every time step $t$ as mandated by the BA Model but was imprecisely defined therein.
	\end{itemize}

We are now in a position to state our first Conjecture of this paper.

\begin{conjec} \label{Cnjctr01}
Given a graph $G_0$ with $m_0 \geq 1$ nodes and an integer $1\leq m \leq m_0$, the PAPA Random Graph $G_t$ generated from it is without any multi-edges or self-loops and is scale-free, i.e., there exists an integer $k_0 > 1$ such that for all $k \geq k_0$ 
it holds that
\begin{equation} \label{sclfr}
	\underset{t \uparrow \infty}{\lim\sup}\ \frac{\|\{v_i \in V_t: d_i(t) = k\}\|}{\|V_t\|} \leq \frac{1}{k^{\gamma}}\ a.s.
\end{equation}
for some $\gamma > 1$.
\end{conjec}

\section{Simulations and Results}

We run our simulations with three different values of $\ep = 0.01, 0.05, 0.1$, the threshold for the strong friendship between a pair of users. In each of the three cases, we start with a complete graph with $m_0 = m$ nodes, where $m$ is the number of nodes that every new node is connected to among the existing set of nodes. Through subsequent addition of nodes done by PAPA as above, we obtain and plot the expected size (number of nodes) of $\LA_t$ in $G^s_t$ against the size (number of nodes) of the graph $G^s_t$ for a given $m_0$ and $\ep$. 
For any graph $G$, $\|G\|$ denotes the cardinality $\|V\|$ of the set $V$ of its nodes.

\begin{figure}
	\centerline{{\resizebox*{3.6in}{1.8in}{\includegraphics{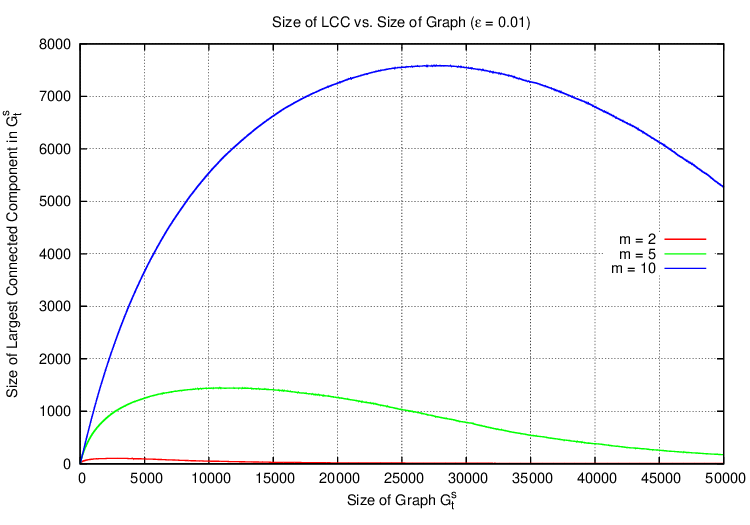}}}}
	\caption{Expected size of the LCC in the strong friendship graph vs. 
		the size of the graph, for $\epsilon = 0.01$}
	\label{fig1}
\end{figure}

\begin{figure}
	\centerline{{\resizebox*{3.6in}{1.8in}{\includegraphics{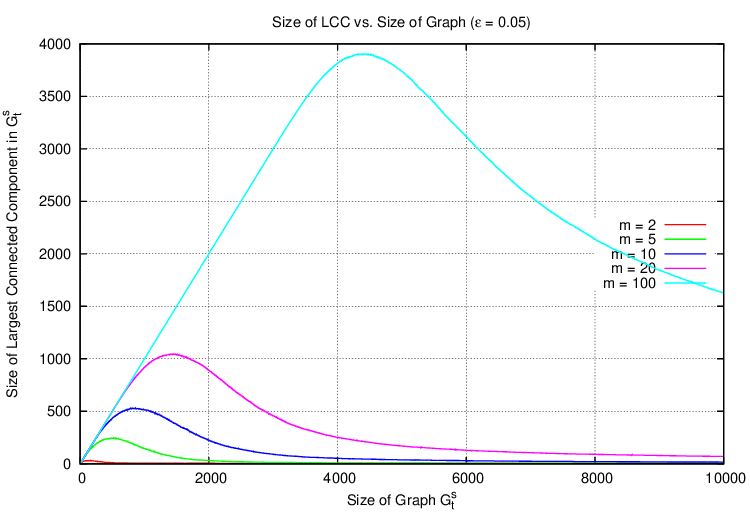}}}}
	\caption{Expected size of the LCC in the strong friendship graph vs. 
		the size of the graph, for $\epsilon = 0.05$}
	\label{fig2}
\end{figure}

\begin{figure}
	\centerline{{\resizebox*{3.6in}{1.8in}{\includegraphics{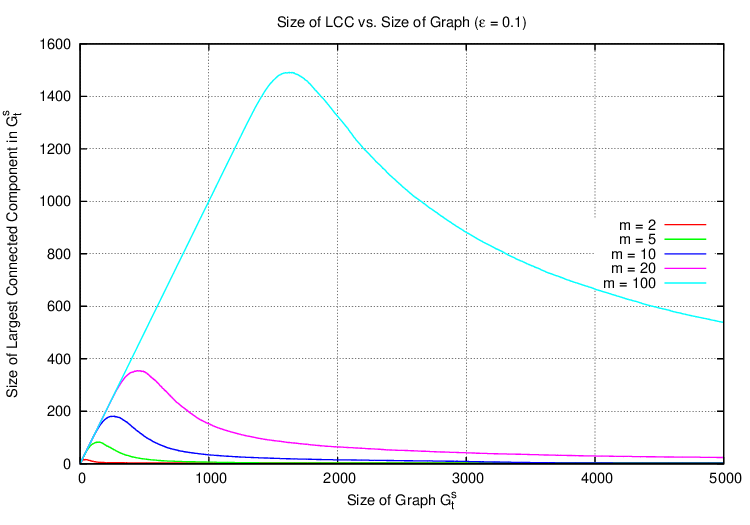}}}}
	\caption{Expected size of the LCC in the strong friendship graph vs. 
		the size of the graph, for $\epsilon = 0.1$}
	\label{fig3}
\end{figure}

We observe from the Figures \ref{fig1}, \ref{fig2} and \ref{fig3} (see after References) that the expected size of the LCC in $G^s_t$ increases till a point (peak) and starts decreasing thereafter in each of the plots. Based on these observations, we now state our second Conjecture of this paper.

\begin{conjec}\label{Cnjctr02}
There exists a threshold real number $0 < \ep_0 < 1$ such that, for all $0 < \ep < \ep_0$, integers $m_0 \geq 1,\ m = m_0$ and a given complete graph $G_0$ with $m_0$ nodes, there exist  finite integers $\tau(\ep, m_0) >  m_0$ and $m_0 \leq \sigma(\ep,m_0) \leq \tau(\ep,m_0)$ satisfying the following property for $\LA_t$ generated by PAPA from $G_0$:
\begin{equation} \label{socnetprop}
E\left[\|\LA_t\|\right] \stackrel{t \uparrow \tau(\ep,m_0) - m_0}{\uparrow} \sigma(\ep,m_0)  \stackrel{t \downarrow \tau(\ep,m_0) - m_0}{\uparrow} E\left[\|\LA_t\|\right].
\end{equation}
\end{conjec}

\noi Note that the above Conjecture \ref{Cnjctr02} automatically implies that
\begin{equation} \label{socnetlim}
\underset{t\geq 0}{\sup}\ E\left[\|\LA_t\|\right] \leq \sigma(\ep,m_0),
\end{equation}
i.e., $E\left[|\LA_t|\right] \leq \sigma(\ep,m_0)$ for all time $t \geq 0$.

\section{Conclusions}

Our observations suggest one possible explanation for the rise and fall of online social networks, i.e.,  the core (LCC) in such a network with increasing number of nodes starts reducing in size after reaching a peak. Since this might be the effect of preferential attachment where a popular user is likely to befriend a large number of other users, it would be interesting to see whether the same observation holds for random  networks that have a restriction on the number of friends that each node can have, i.e., where we only 
allow `preferential attachment with limit' or `limited rich-gets-richer' connections. 
Another interesting open problem might be to check if the size of the LCC increases monotonically with that of the underlying network if $\ep > \ep_0 $ where $\ep_0$ is as in Conjecture \ref{Cnjctr02} above. Estimating explicit bounds for $\tau(\ep, m_0),\sigma(\ep,m_0)$ in terms of these network parameters might also be looked into. Also, studying the same problem for $G_0$ to be a non-complete graph and/or $1 \leq m < m_0$ will be interesting. We believe that Conjecture \ref{Cnjctr02} should hold in such situations.


\newpage


\small
\begin{thebibliography}{2200}

\bibitem{bailey}
N. Bailey. Mathematical Theory of Infectious Diseases. Mathematics in 
Medicine Series. Oxford University Press, 1987.

\bibitem{barabasi}
A. L. Barabasi and R. Albert. Emergence of scaling in random networks. Science 286 (5439): 509-512, 1999. 

\bibitem{berger}
N. Berger, C. Borgs, J. T. Chayes and A. Saberi. Asymptotic behavior and distributional limits of preferential attachment graphs.
The Annals of Probability 42 (1): 1-40, 2014.

\bibitem{cannarella}
J. Cannarella and J. A. Spechler. Epidemiological modeling of online social network dynamics. 
arXiv:1401.4208, 2014.

\bibitem{dunbar}
R. I. M. Dunbar. Neocortex size as a constraint on group size in primates. 
Journal of Human Evolution 22 (6): 469-493, 1992. 

\bibitem{easley}
D. Easley and J. Kleinberg. Networks, Crowds, and Markets:
reasoning about a highly connected world. Cambridge University Press, Cambridge UK, 2010.

\bibitem{garavaglia}
A. Garavaglia. Preferential Attachment Models for Dynamic Networks. Ph.D. Thesis, TU Eindhoven, 2019.

\bibitem{gillette}
F. Gillette. The rise and inglorious fall of myspace. Businessweek magazine, 2011. 

\bibitem{HFSTD01}
R.\ van der Hofstad.\ Random Graphs and Complex Networks.\ Cambridge Univ.\ Press, Cambridge UK, 2018.

\bibitem{huberman}
B. A. Huberman, D. M. Romero and F. Wu. Social networks that matter: Twitter under the microscope,
First Monday 14 (1), January 2009.

\bibitem{marlow}
C. Marlow, L. Bryon, T. Lento and I. Rosenn.
Maintained relationships on Facebook, http://overstated.net/2009/03/09/maintained-relationships-on-facebook, 2009.

\bibitem{onnela}
J. P. Onnela, J. Saramaki, J. Hyvonen, G. Szabo, D. Lazer, K. Kaski, J. Kertesz and A. L.
Barabasi. Structure and tie strengths in mobile communication networks. Proc. Natl. Acad. Sci. USA, 104: 7332-7336, 2007.

\bibitem{wu}
S. Wu, A. Das Sarma, A. Fabrikant, S. Lattanzi and A. Tomkins. Arrival and departure dynamics
in social networks. Proceedings of the Sixth ACM International Conference on Web Search
and Data Mining (WSDM), 233-242, 2013.

\end{thebibliography}
\end{document}